# Peculiar magnetic properties of NC$_6$ and NC$_{12}$ layered compounds from first principles.

Samir F. Matar*,[1]

Lebanese German University (LGU), Sahel-Alma Campus, Jounieh, Lebanon.

*Email: s.matar@lgu.edu.lb; abouliess@gmail.com

[1]formerly at CNRS, University of Bordeaux, ICMCB. 33600 Pessac. France.

**Abstract**.

In the context of characterizing nitrogen poor carbo-nitrides for different applications, identification of an unusual onset of spin polarization of N(p) states has been shown. A full saturation up to 3 $\mu_B$ is demonstrated in extended two-dimensional carbon networks of C$_6$N and C$_{12}$N hexagonal structures refined based on density functional theory calculations. From establishing the energy-volume equations of states in both compounds assuming spin degenerate (non spin polarized) and spin-polarized configurations, the ground state is identified as ferromagnetic. The variation of magnetization with volume points to strongly ferromagnetic behavior.

## I- Introduction:

The onset of magnetic polarization requires a significant localization of the states likely to carry a finite magnetic moment. Such localization is illustrated by a high density of states (DOS) at the Fermi level $n(E_F)$ in an initially spin degenerate electron system configuration. $n(E_F)$ can be quantified from calculations within density functional theory DFT [1,2] and inferred from the Stoner theory of band (ferro)magnetism [3] whereby localization leading to a large $n(E_F)$ magnitude is an indication of unstable electronic system in spin degenerate configuration (also labeled non spin polarized NSP) and that it should stabilize by spin polarization (SP) via dispatching the electrons into two spin populations: majority spins ↑ and minority spins ↓. The difference between ↑ and ↓ spin populations gives a finite magnetic moment. In general magnetization develops on d states of transition metals or f states of rare-earths and actinides: $n$f ($n$= 4, 5 respectively). Regarding transition metals $n$d ($n$ =3, 4, 5), the first period developing finite magnetization are ferromagnetic metals Fe, Co and Ni, but not metals of 2$^{nd}$ and 3$^{rd}$ periods because their 4d and 5d bands are too broad to allow for d states to localize enough. Also rare earth Gadolinium is a ferromagnet at room temperature with M= 7$\mu_B$ magnetic moment. Note that while in 3d ferromagnetic metals the onset of the magnetic moment is of interband nature (mediated by the electron gas); in Gd the moment arises from 4f intraband spin polarization.

On the other side the elements devoid of d states have p valence external subshells characterized by large spatial expansion, which leaves them little chance for the development of magnetic moment. However ordered magnetic moments were identified in hexaborides $AE$B$_6$ ($AE$ = Ca, Sr) [4] as well as in CdS doped with main group elements [5]. Also the onset of finite magnetization carried by N-p states thanks to the localization of N(p$_z$) was recently shown, based on computations within DFT in a layered carbon graphitic-like AlB$_2$-type C$_2$N structure with 1.1 $\mu_B$ aligned along $c$-hexagonal axis [6]. The AlB$_2$-type C$_2$N structure (cf. Fig. 1) can also be viewed as a honeycomb arrangement of C sublattice interlayerd by N, a feature close to the lithium intercalation compounds [7] used as electrodes in electrochemical processes of battery charge-discharge. In this context two graphitic anode compositions were identified: $LiC_{12}$ and $LiC_6$ [8] and characterized by *in situ* neutron diffraction with layered-like structure within *P6/mmm* space group (like $AlB_2$-type $C_2N$), cf. Fig. 1c) and 1d).

Following the unusual results of magnetization of main group elements in certain structural and electronic conditions [4, 5] and of N(p) in layered $AlB_2$-type $C_2N$ *P6/mmm* [6] on one hand and the context of extended two dimensional (2D) carbon network in $LiC_6$ and $LiC_{12}$ [8] in the same space group as $AlB_2$–$C_2N$ (cf. Table 1), it became necessary to further investigate the magnetic behavior of N in such extended 2D carbon networks of $C_6N$ and $C_{12}N$ (Fig. 1) obtained from geometry optimizations starting from the corresponding Li–based experimental structures.

Also this study of $C_6N$ and $C_{12}N$ is inscribed to a certain extent in the context of early investigations more than 25 years ago of nitrogen poor carbonitrides with CVD/PVD growth experiments aiming at preparing firstly layered then ultra hard compounds as $C_{11}N_4$ for applications as coating materials in tooling machinery [9, 10].

The paper reports on such original investigations and shows that the effect of carbon lattice expansion leads to the full polarization of all three N(p) electrons with 3 $\mu_B$ saturated magnetic moment in a ferromagnetic ground state from energy-volume equations of states, and to different magnetovolume behaviors with weak and strong ferromagnetic behaviors of $NC_6$ and $NC_{12}$ respectively.

## II- Computation framework

Within DFT plane waves VASP code [11, 12] to geometry optimize atomic positions and lattice parameters leading to ground state configuration with minimized inter-atomic forces. Two sets of calculations considering spin degenerate (non spin polarized NSP) and spin polarized (SP) configurations for each one of the two carbo-nitrides $C_6N$ and $C_{12}N$. The projector augmented wave (PAW) method [12,13] with potentials built within the generalized gradient approximation (GGA) for an account of the effects of exchange and correlation [14]. Within our computational scheme the conjugate-gradient algorithm [15] was used to relax the atom positions of the different chemical systems into their ground state structure. The structural parameters were considered to be fully relaxed when forces on the atoms were less than 0.02 eV/Å and all stress components below 0.003 eV/Å$^3$. The tetrahedron method with Blöchl corrections [16] was applied for both geometry relaxation and total energy calculations. Brillouin-zone (BZ) integrals were approximated using the special k-point sampling of Monkhorst and Pack [17]. The calculations were converged at an energy cut-off of 500 eV for both compounds. The **k**-point integration is carried out with a starting mesh of 6 × 6 × 6 up to 10 × 10 × 10 for best convergence and relaxation to zero strains.

Properties relevant to electron localization are obtained from real space analysis of electron localization function (ELF) according to Becke and Edgecomb [18] which was initially formulated for Hartree–Fock approach then adapted to DFT methods. ELF is based on the kinetic energy in which the Pauli Exclusion Principle is accounted for: ELF = $(1+ \chi_\sigma^2)^{-1}$ with $0 \leq ELF \leq 1$, meaning that ELF is a normalized function. In this expression the ratio $\chi_\sigma = D_\sigma/D_\sigma^0$, where $D_\sigma = \tau_\sigma - \frac{1}{4}(\nabla\rho_\sigma)^2/\rho_\sigma$ and $D_\sigma^0 = 3/5 (6\pi^2)^{2/3}\rho_\sigma^{5/3}$ correspond respectively to a measure of Pauli repulsion ($D_\sigma$) of the actual system and to the free electron gas repulsion ($D_\sigma^0$) and $\tau_\sigma$ is the kinetic energy density. In this paper we use ELF planes along selected orientations of the cell to show differentiated electron localizations with color maps: bleu areas for no localization, red for full localization and green for free electron like localization.

## III- Geometry optimization and energy dependent results

Geometry optimizations pertain to atomic relaxations with no structural constrains. Starting from the lithium-carbon compounds iterative calculations led to minimize the inter-atomic forces while keeping the symmetry in *P6/mmm* space group. The protocol was carried out for both spin degenerate (NSP) and spin polarized (SP) configurations for each composition.

III-1 Electron localization function maps.

An illustration of the results after full geometry relaxation is obtained from the electron localization ELF mapping. Figures 2 a) and b) show the basal plane ELF projection of N contours in $C_6N$ (top) and $C_{12}N$ (bottom) for four adjacent cells. The color scale shown at the bottom of the figures is from 0 (blue) to 1 (red). The presence of large blue areas between the strong concentrations of electrons around N indicate localization and isolation of individual N. Vertical ELF contour plot at Figs. 2c) and 2d) show the electron concentration around C-C pairs with 1 layer in $C_6N$ and two layers in $C_{12}N$ on one hand and the localization around N along the hexagonal c-axis. Non vanishing green free-electron like areas indicate bonding between N and C. Lastly in Figs. 2 e) and f) show for $C_6N$ and $C_{12}N$ the strong localization of electron between C-C in honeycomb like carbon network.

III-2 Energy related results analyses.

Table 1 provides the starting crystal data of $LiC_6$ and $LiC_{12}$ used to optimize the corresponding carbonitrides in both spin configurations, NSP and SP. Also for the sake of completeness the data of $C_2N$ are included at first column. The *x* internal coordinate of carbon as well as the C-C distances change little from Li to N cases meaning that the carbon host network undergoes negligible changes (even for $C_2N$ with a single C-C pair and where d(C-C) is only slightly smaller). *a* lattice parameter changes little from $C_6N$ to $C_{12}N$ on one hand and from NSP to SP magnetic configuration on the hand. This implies that the *a* parameter defining the horizontal planes is hexagonal lattice, is controlled by the carbon network.

Using the NSP calculations trends of cohesive energies $E_{coh}$ were established within the $C_xN$ series averaged as per one atom for better comparison. N and C energies were calculated based on the atom placed in a cubic box. Then E(N)= -6.830 eV and E(C) = -7.11 eV. From the energy optimization, $E_{Tot.}(C_6N)$ = -56.44 eV and $E_{Tot.}(C_{12}N)$ = -111.446 eV while $E_{Tot.}(C_2N)$ = -21.52 eV. Then the cohesive energies are averaged as per one atomic constituent to enable trends. They amount to (in eV/at.): $E_{coh.}(C_2N)$ = -0.16 eV/at.; $E_{coh.}(C_6N)$ = -0.99 eV/at.; $E_{coh.}(C_{12}N)$ = -1.48 /at. The increase of cohesive energy is in line with the extension of the carbon network and while it is almost 6 times from $C_2N$ to $C_6N$, it amounts to 33% from $C_6N$ to $C_{12}N$ which is most cohesive.

Major changes are observed from NSP to SP for the *c* hexagonal parameter. This was equally observed for $C_2N$ (1$^{st}$ column). The outcome is that the volume of the cell is larger in the spin polarized configuration as a consequence of the onset of magnetization on nitrogen which amounts to M= 3 $\mu_B$ in both $C_6N$ to $C_{12}N$ compounds. Note that in $C_2N$ M= 1.1 $\mu_B$.

One first observation that can be proposed is that $M(C_2N)$ = 1.1 $\mu_B$ is not a saturated magnetization and that saturation requires an extended carbon network such as the one in $C_6N$ to $C_{12}N$ where it amounts to 3 $\mu_B$ i.e. with the polarization of all three p electrons. Somehow this meets with the situation of rare-earth *RE* Gd where the 7 $\mu_B$ magnetic moment arises from the seven electrons of the 4f-half filled subshell. With this parallel approach it can be suggested that alike *RE* p-magnetization here is equally of intraband polarization nature. We shall further develop on this original observation in next sections.

III-3 Spin-degenerate density of states DOS and origin of the magnetic instability.

The origin of the spin polarization should be assessed based on the projection of the spin degenerate NSP DOS. Fig. 3 top panel shows the site projected DOS for $C_6N$ exemplarily. The DOS of N and all six carbons are projected. The energy reference along *x*-axis is with respect to the Fermi level $E_F$ which is crossed by a large N-DOS whereas C shows little contribution with nevertheless a small peak following the shape of N-DOS whence the quantum mixing between C and N leading to the chemical bonding. As stated in the introduction, such large $n(E_F)$ magnitude which is of *p*-character (*s*-states are far lower in energy) is an indication of unstable electronic system in spin degenerate state The role of each orbital is shown by the decomposition over $p_x$, $p_y$ and $p_z$ in the lower panel of Fig. 3. Two kinds of p-DOS can be seen, two degenerate in plane $p_x$ and $p_y$, broader (notice the DOS-shoulder) than out-of-plane $p_z$ which resembles more the carbon DOS at $E_F$. These results mirror the ELF projections in Fig. 2 where panels a) and b) show the $p_x$ and $p_y$ ELF isolated from carbon whereas panels c) and d) exhibit finite ELF between N and C-C pairs whence the bonding. Then all three N(p) contribute to the instability which eventually leads to magnetic polarization. As a matter of fact subsequent SP calculations lead to the onset of magnetization of 3 $\mu_B$ in both carbonitrides. The illustration of such integer magnetization identified in both compounds is illustrated in Figure 4 by the spin (UP ↑ and DOWN ↓) projected total DOS where the Fermi level is now in a small gap in $C_6N$ and slightly larger but well defined in $C_{12}N$, implying that the three p electrons are fully polarized as UP ↑ spins. These are called majority spins because the corresponding DOS are shifted to lower energy versus DOWN ↓ (minority spins) shifted to higher energy as can be seen in the SP DOS plots. Consequently starting from the high DOS at $E_F$ in the NSP calculations (Fig. 3), the electron system relaxes to a SP ground by reducing the large $n(E_F)$ magnitude. The two compounds are then predicted as ferromagnetic semi-conductors. This behavior is close to that observed for $C_2O$ and $CrO_2$ [19] strong half metallic ferromagnets as well as $Co_3Sn_2S_2$ [20, 21]. .

III-4 Energy-volume equations of state and volume dependent magnetic behavior.

At this point the evaluation of the impact of volume changes upon the onset and change of magnetization requires the assessment of the energy–volume equation of states (EOS) of each compound in its two magnetic configurations NSP/SP with calculations around minima found from geometry optimization (Table 1). The resulting curves are plotted in Fig. 5. In both compounds the ground state configuration is magnetic (SP) with large stabilization versus NSP: $\Delta E_{C_6N}$ (SP-NSP) = -1.91 eV/cell and $\Delta E_{C_{12}N}$(SP-NSP) = -2.19 eV/cell and the stabilization for $C_{12}N$ is much larger.

The fit of the curves which show a quadratic behavior is done with 3rd order Birch EOS [22]:

$$E(V) = E_o(V_o) + [9/8]V_oB_o[([(V_o)/V])^{2/3}-1]^2 + [9/16]B_o(B'-4)V_o[([(V_o)/V])^{2/3}-1]^3$$

Where $E_o$, $V_o$, $B_o$ and $B'$, the fit parameters are respectively the equilibrium energy, the volume, the bulk modulus and its pressure derivative. The obtained corresponding values are given in the insert. $\chi^2$ is the goodness of fit indicator.

One first result is the difference of magnitudes between bulk modules, larger for $C_{12}N$ for both NSP and SP on one hand and the decrease of $B_0$(SP) versus $B_0$(NSP) accompanying the reverse trends of volume; i.e. the larger the volume, the smaller the bulk modulus. The SP/NSP crossovers are at 70 and 120 Å$^3$ volume magnitudes for $C_6N$ and $C_{12}N$.

The results can be further assessed by plotting the magnetization as function of volume. The scattered points shown in Figure 6 present a different behavior at low volume, i.e. with a progressive increase in $C_6N$ and a much steeper increase in $C_{12}N$ both tending to 3 $\mu_B$, which is in both carbonitrides the saturation magnetization. Note however that saturation is reached in $C_{12}N$ before the NSP/SP crossover line oppositely to $C_6N$ where magnetization collapses right at the vertical red line of crossover. Then it can be suggested that $C_{12}N$ is a stronger ferromagnet than $C_6N$.

IV- Conclusion.

In spite of the focus of present work on the magnetic properties of binary carbon rich $C_xN$ with results that encourage synthesis endeavor, the topic is yet inscribed to a certain extent in the context of early investigations more than 25 years ago of nitrogen poor carbonitrides in a European Network for the search of new Ultra-hard Materials. The know-how-to in synthesizing such compositions with modern

CVD/PVD and other growth techniques leading for instance to original compositions as $C_{11}N_4$ carbonitride in both 2D and 3D forms [9,10] may cast confidence that further growth experiments with smaller amounts of nitrogen might lead to new nitrogen deficient carbonitrides with compositions ranging from $C_6N$ to $C_{12}N$ which can be then proposed to the physics community for further characterization, especially the unusual magnetic behavior.

Table 1. Layered NC$_x$ (x= 2, 6, 12). Space group *P*6/*mmm*; N°191. Distances are in units of Å
(1Å = 10$^{-10}$m)

| Calculated [6] NSP (SP) NC$_2$ (AlB$_2$-type) | Experimental [8] LiC$_6$ | Calculated NSP (SP) NC$_6$ | Experimental [8] LiC$_{12}$ | Calculated NSP (SP) NC$_{12}$ |
|---|---|---|---|---|
| N (1*a*) 0,0,0 | Li (1*a*) 0,0,0 | N (1*a*) 0,0,0 | Li (1*a*) 0,0,0 | N (1*a*) 0,0,0 |
| *a* = 2.41 (2.41) | *a* = 3.68 | *a* = 4.27 (4.26) | *a* = 4.268 | *a* = 4.265 (4.265) |
| *c* = 4.85 (5.32) | *c* = 4.30 | *c* = 4.42 (5.24) | *c* = 7. 022 | *c* = 8.078 (8.781) |
| C (2*d*) 1/3,2/3, ½ | C (6*k*) 0.333,0,½ | C(6*k*) 0.334,0,½ | C(12*n*) 0.333,0,¼ | C(12*n*) 0.336,0,¼ |
| d(N-C) = 3.0<br>d(C-C) =1.39 | d(Li-C) =2.33<br>d(C-C)= 1.44 | d(N-C) = 2.90 (2.95)<br>d(C-C) =1.42 (1.41) | d(Li-C) =2.26<br>d(C-C) = 1.43 | d(N-C) =2.46 (2.62)<br>d(C-C)= 1.42 (1.40) |

**Figures captions**

Figure 1: $C_xN$ layered structures in *P6/mmm* space group shown in simple cell and multiple cells projections to highlight the extended basal plane projection;: a) $C_2N$ [2] b) geometry optimized $C_6N$ and c) $C_{12}N$ based on $LiC_6$ and $LiC_{12}$ experimental structures [3].

Figure 2: Electron localization function ELF slices of title compounds.

Figure 3: NSP site projected DOS in $C_6N$ and N(p) states projected over the x, y and z components

Figure 4: Energy-volume curves of $C_6N$ and $C_{12}N$ in non spin polarized (NSP) and magnetically ordered (spin polarized SP) configurations. In both the ground state is SP (ferromagnetic) at larger volume than NSP.

Figure 5: Spin projected total DOS of ferromagnetic $C_6N$ and $C_{12}N$.

Figure 6: Variation of the magnetization with cell volume in $C_6N$ and $C_{12}N$.

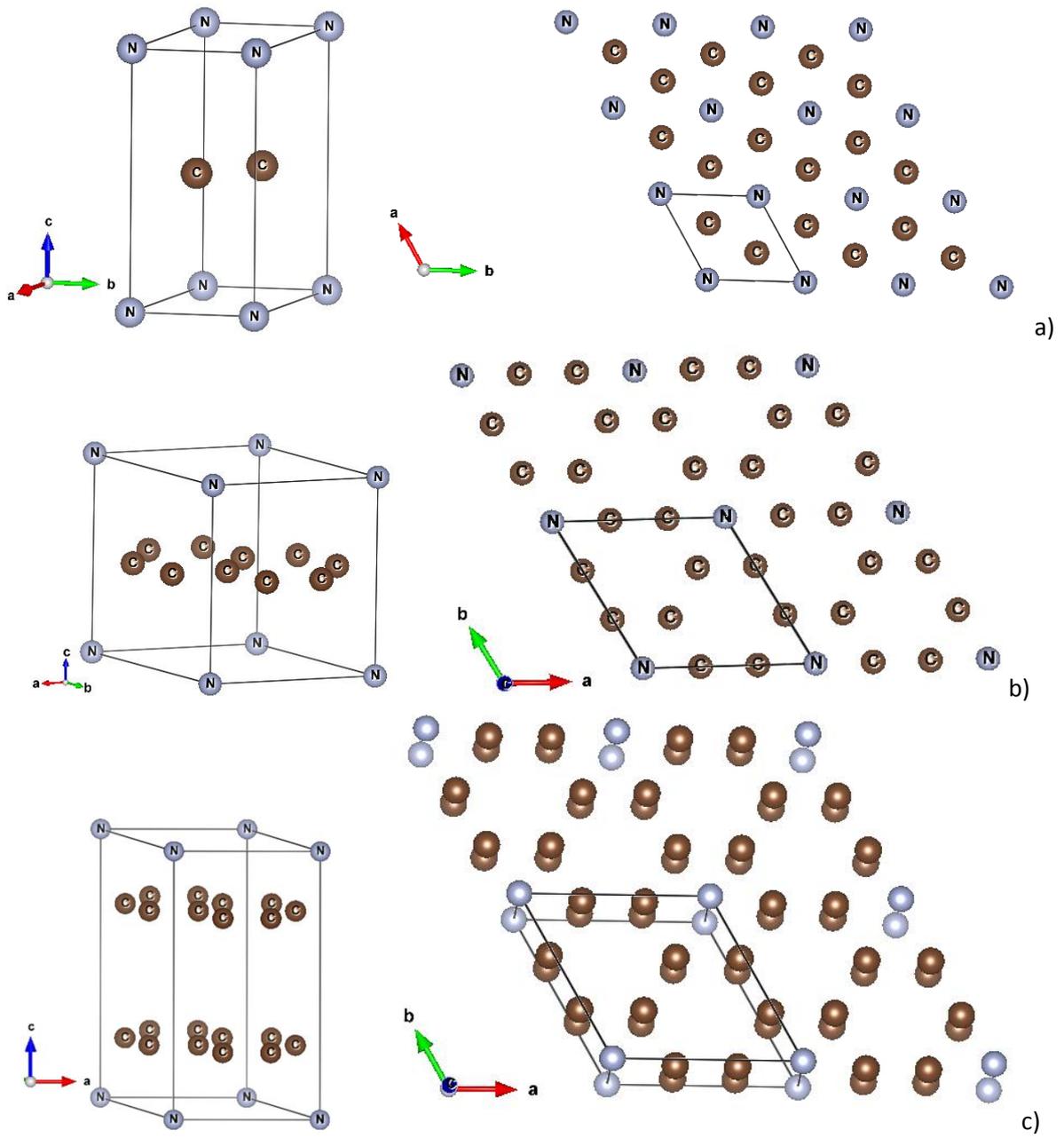

Fig. 1

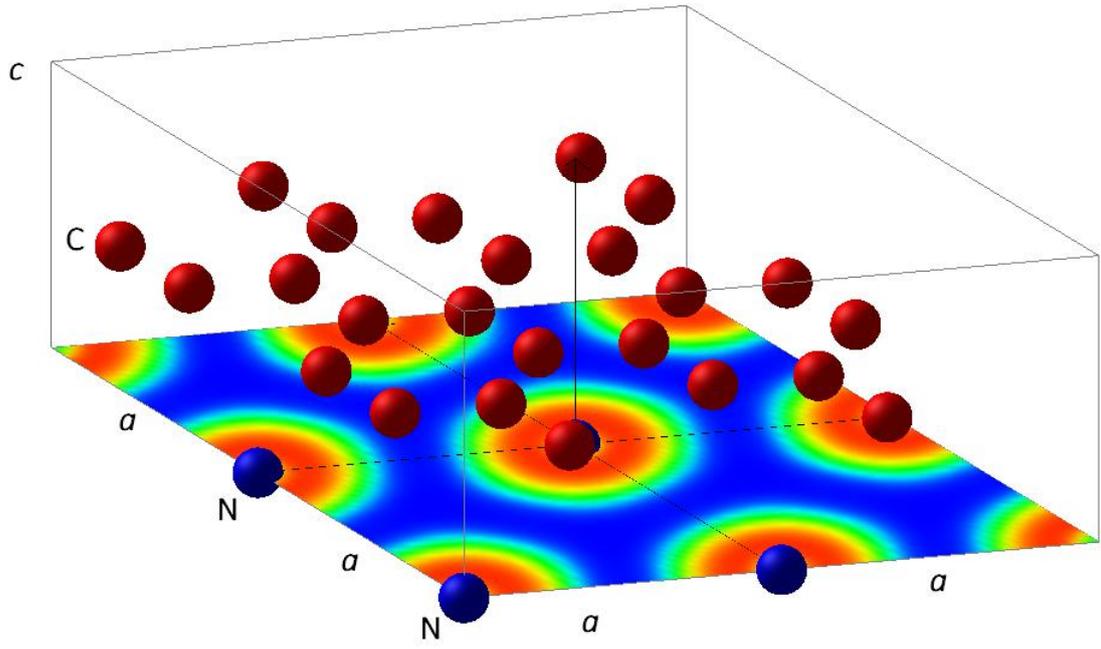

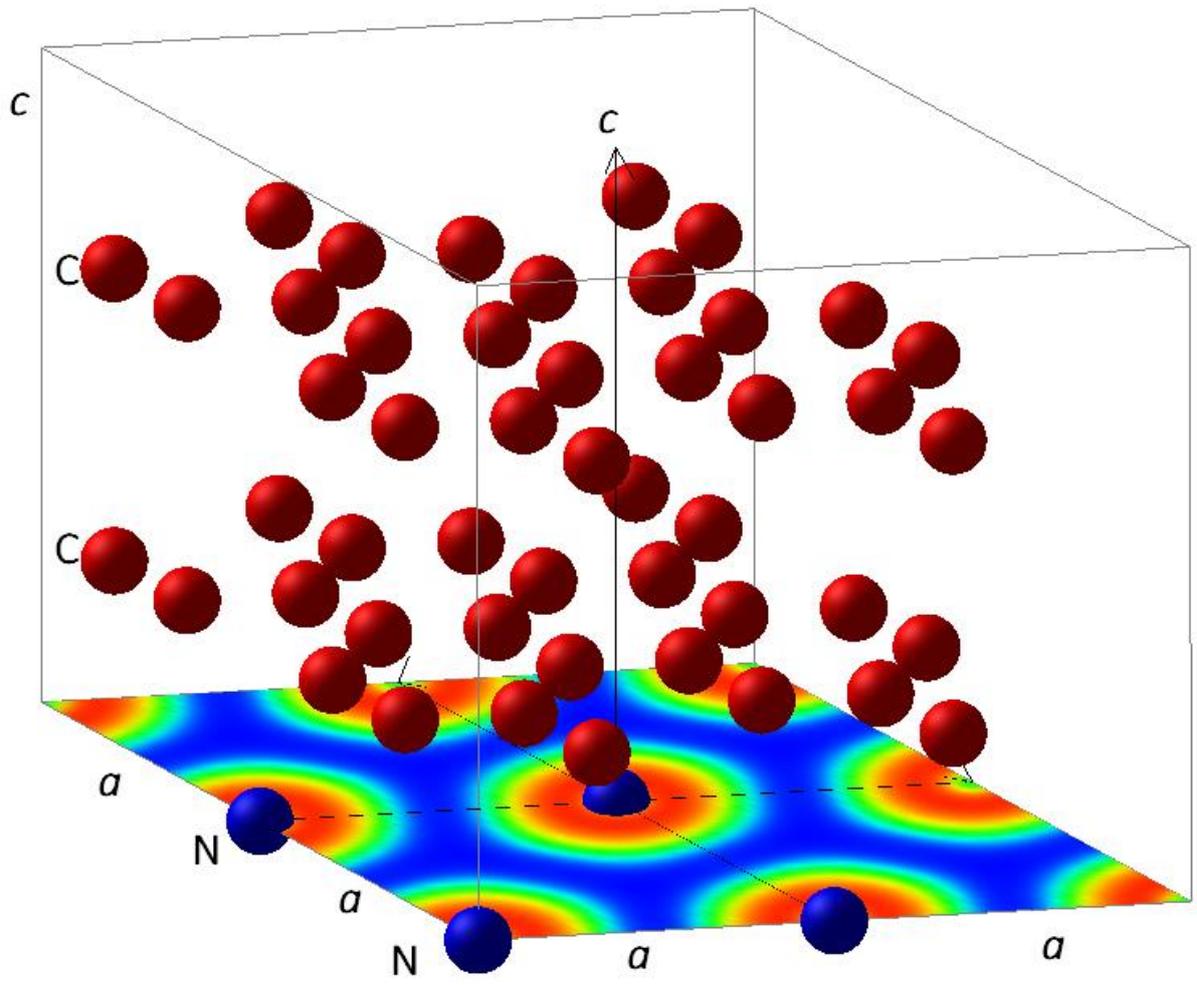

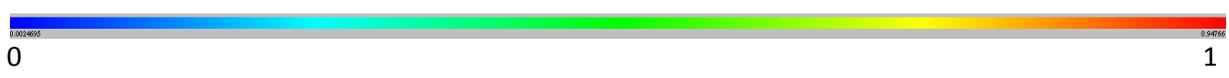

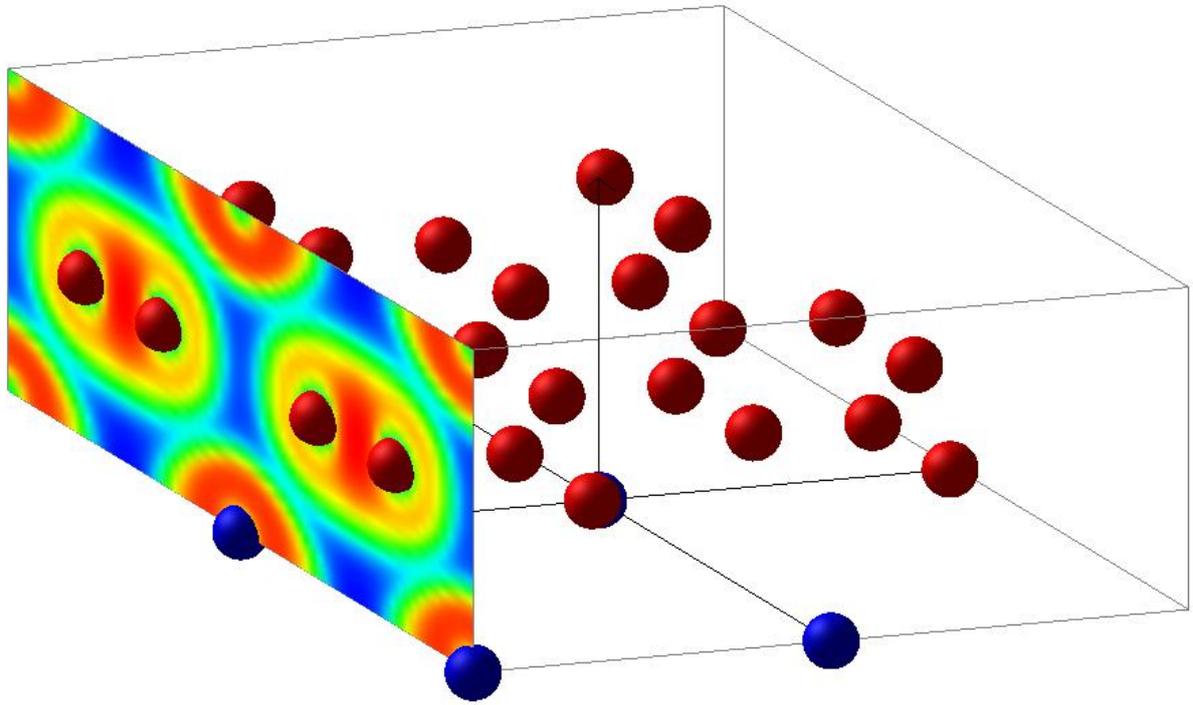

c)

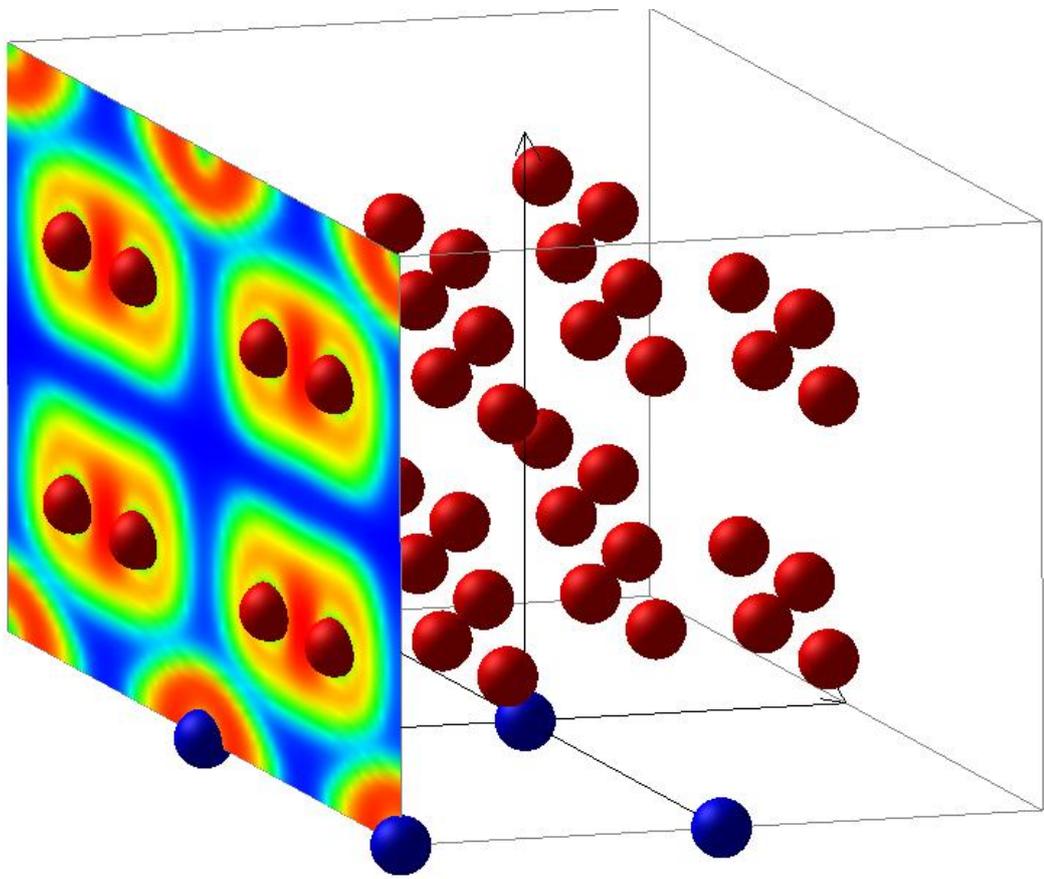

d)

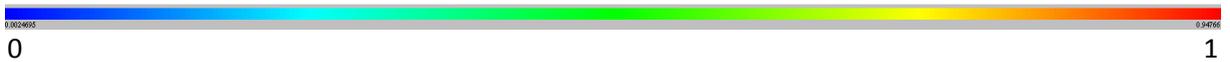

0                                                                                                                                    1

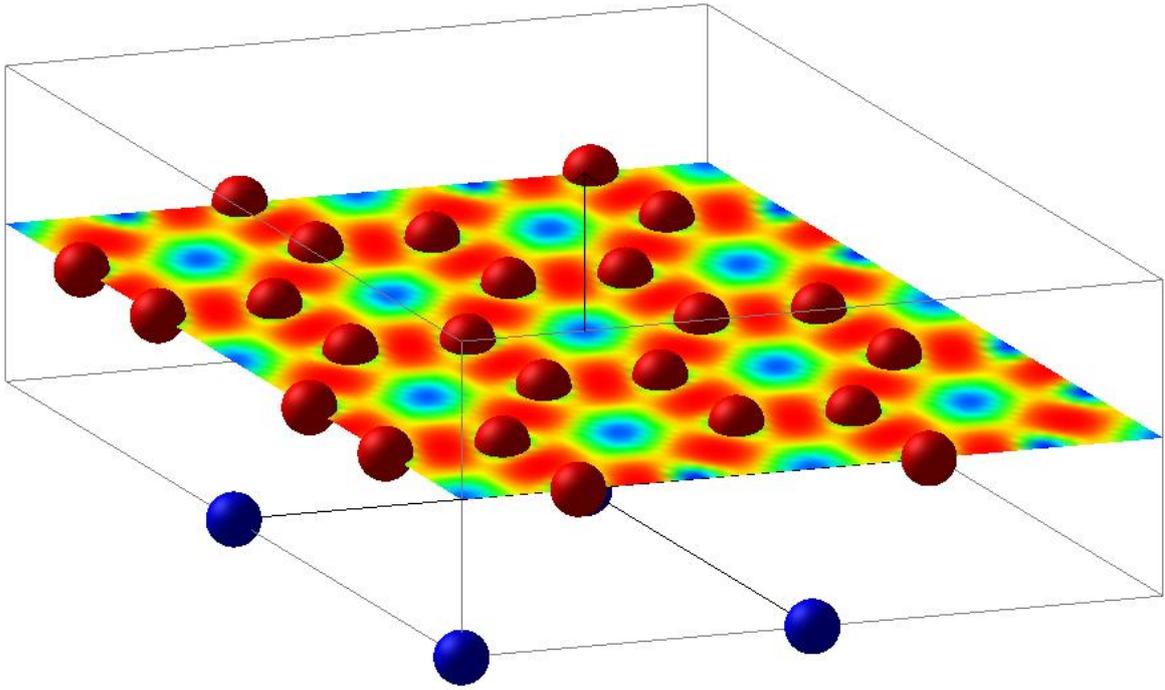

e)

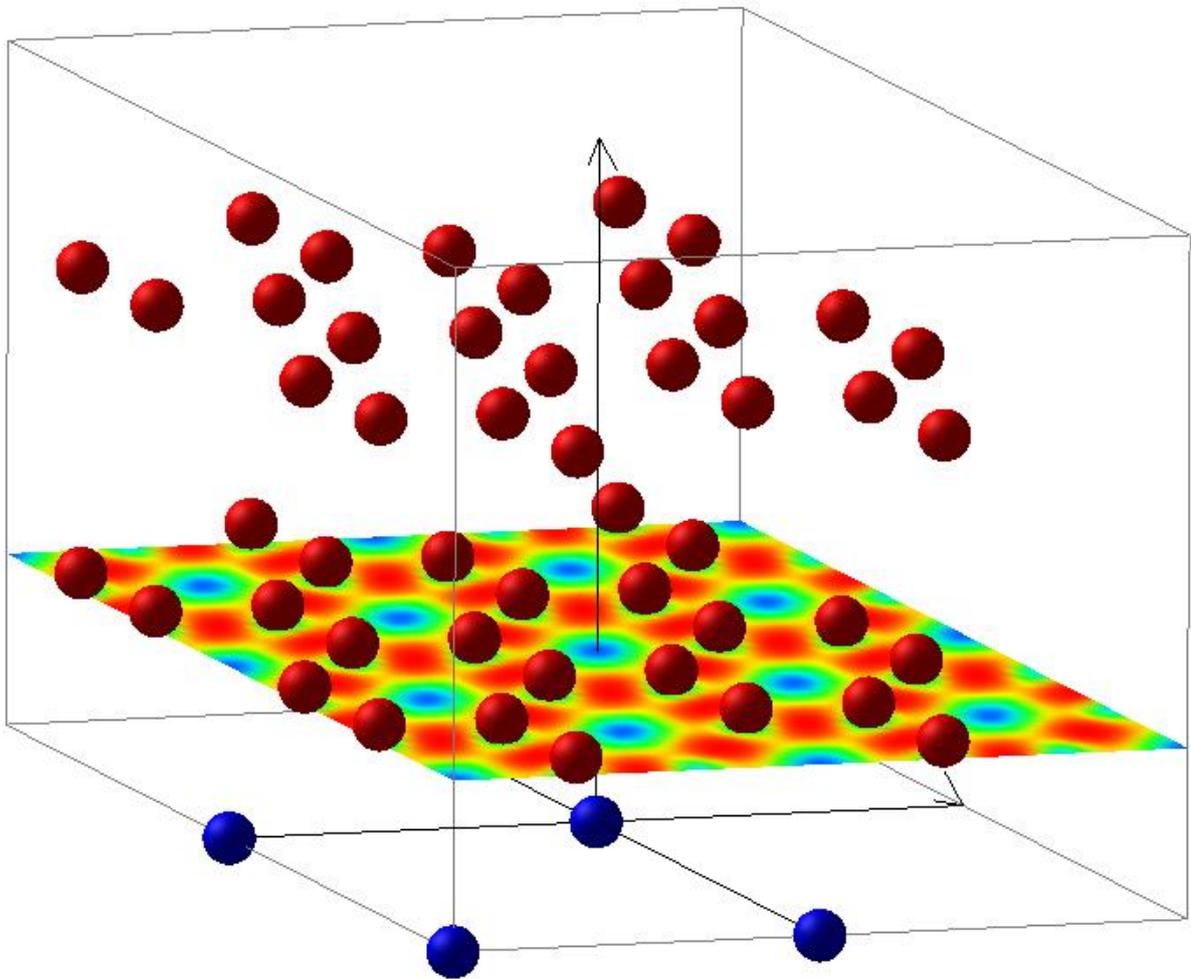

f)

Figure 2

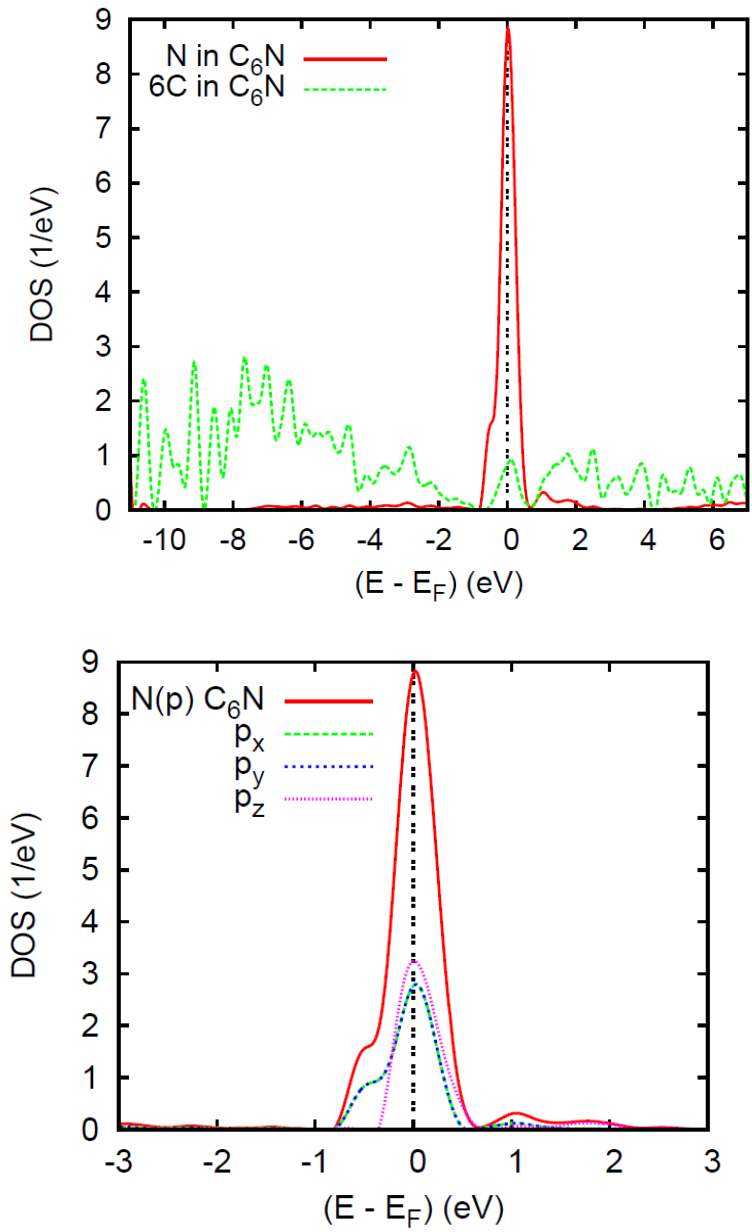

Figure 3

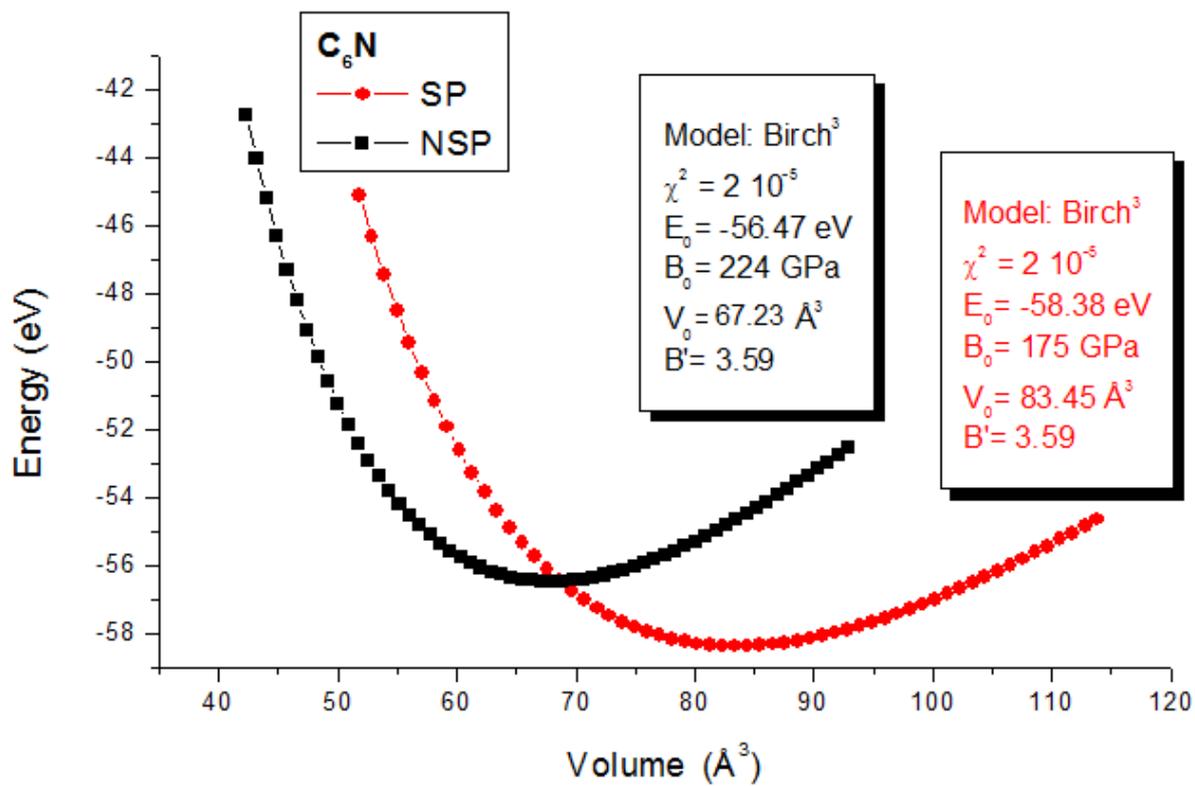

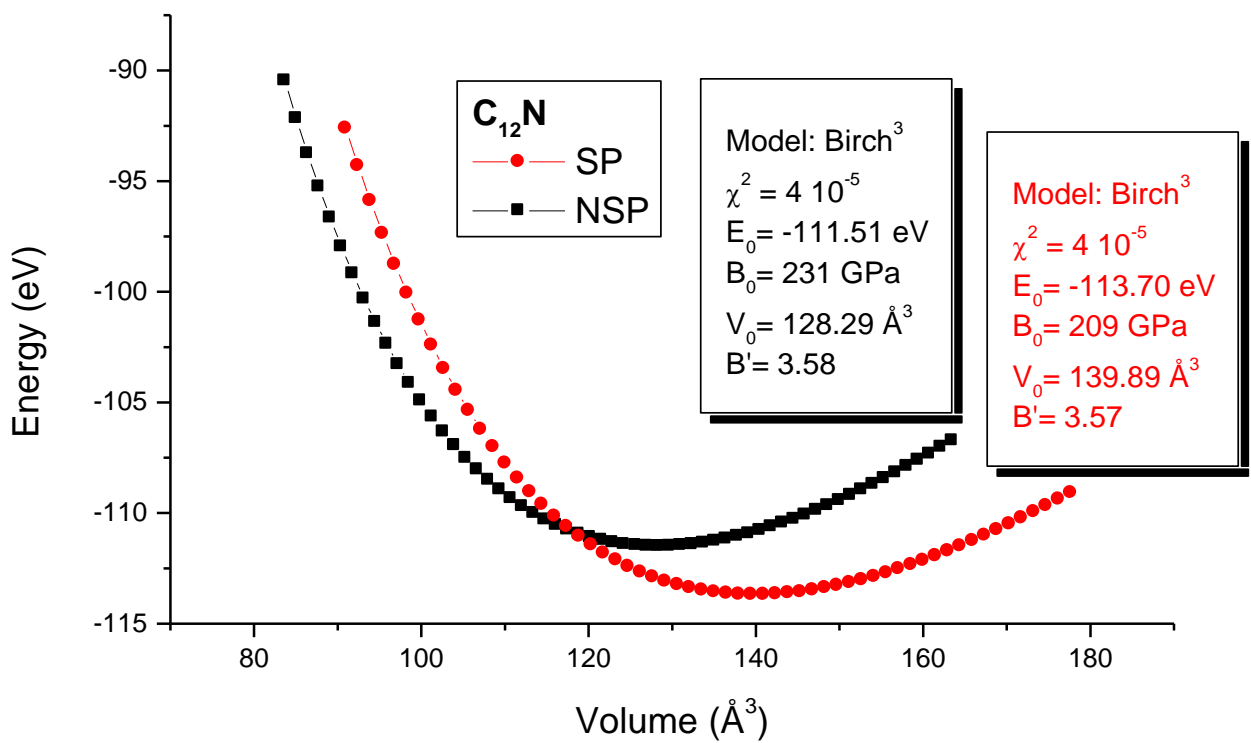

Figure 4.

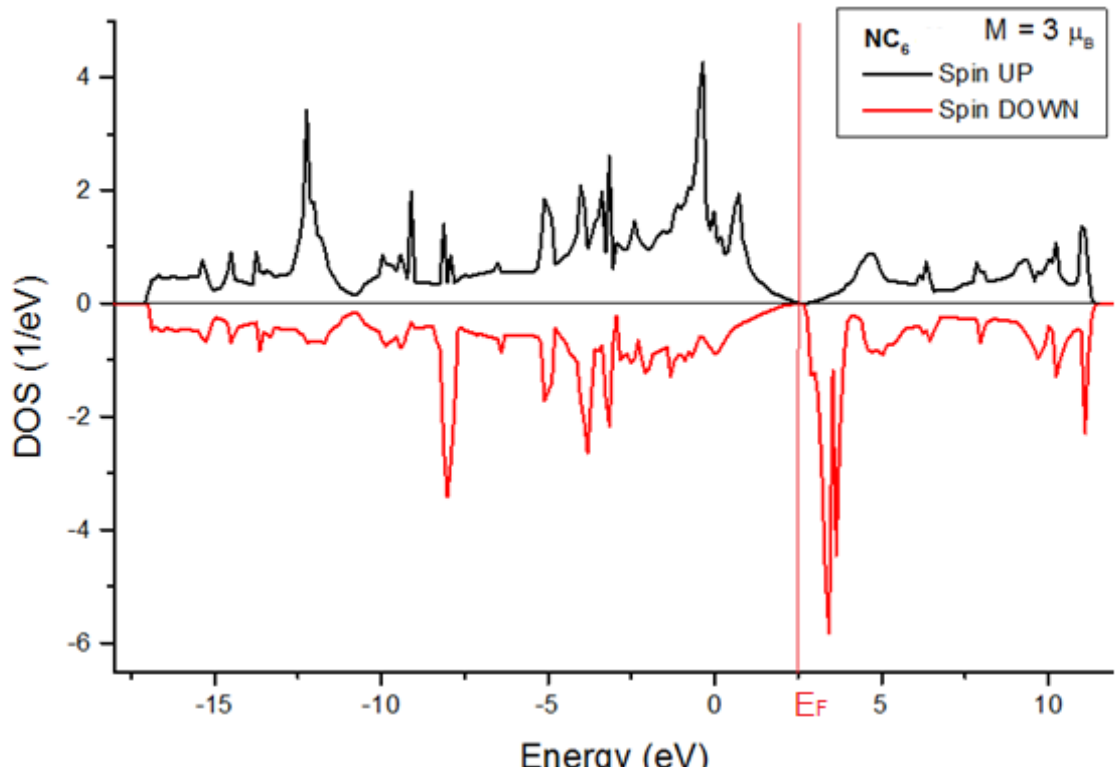

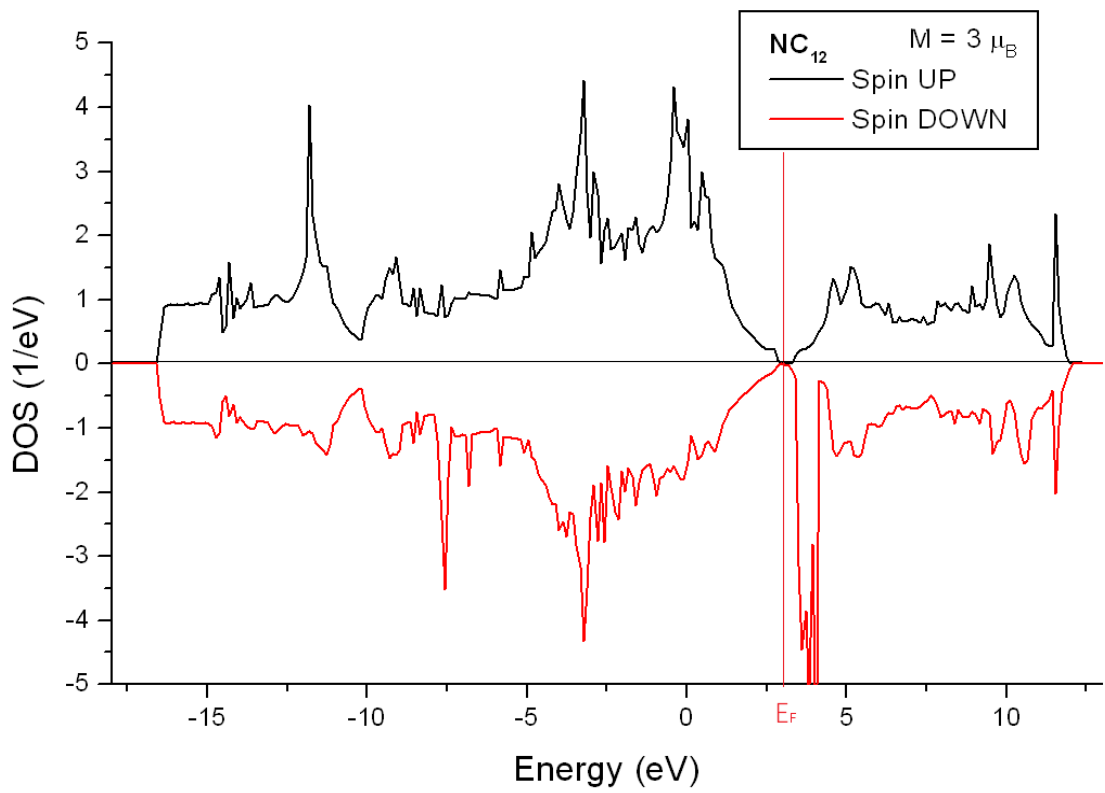

Figure 5

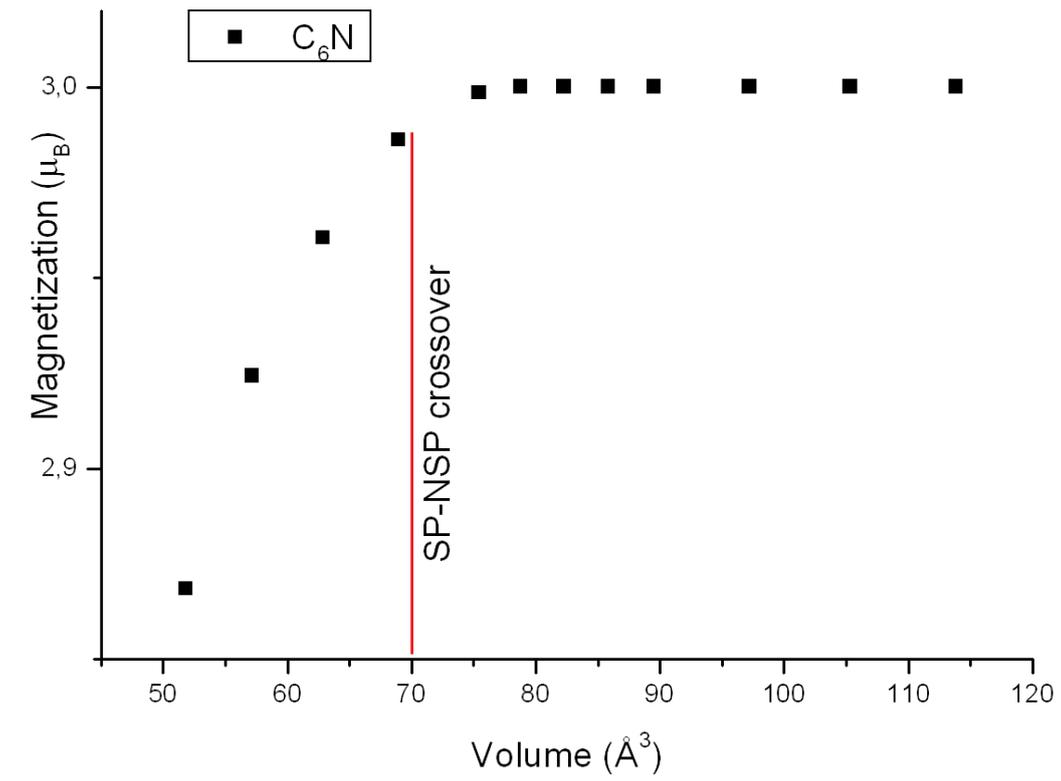
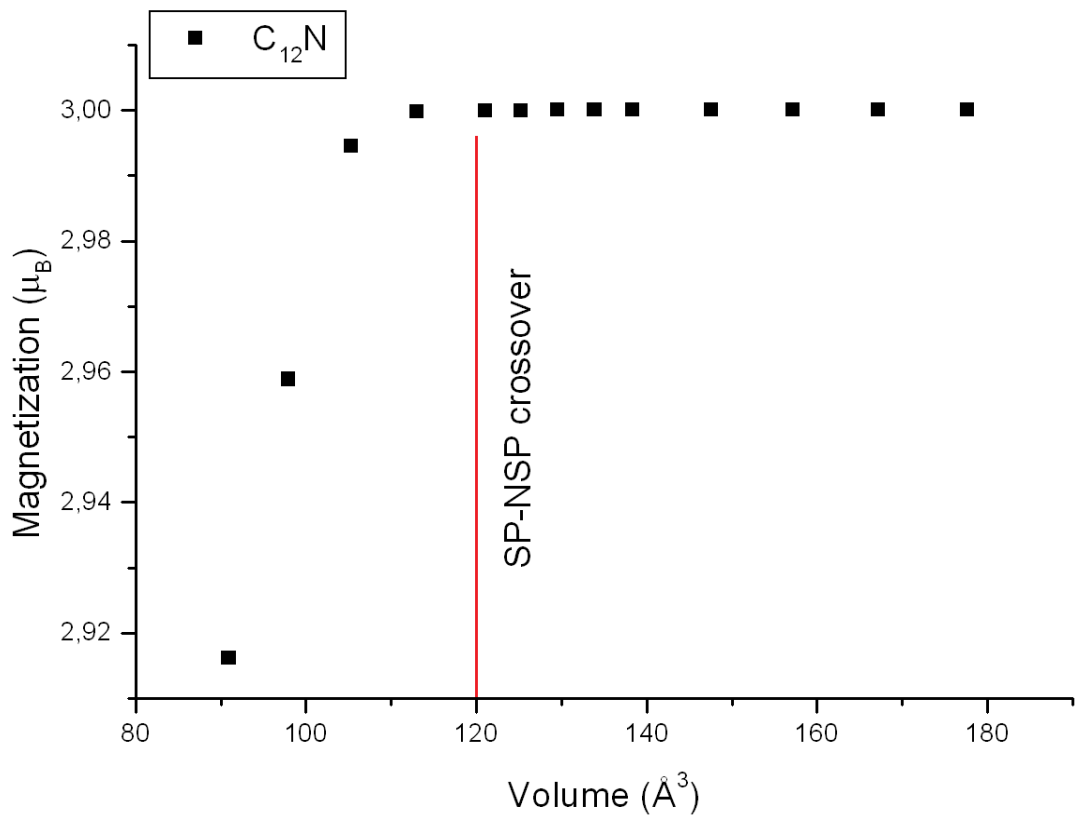

Figure 6